\documentclass[prl,twocolumn,aps]{revtex4}
\usepackage{graphicx}
\usepackage{tikz}
\usepackage{verbatim}

\newcommand{\be}{\begin{equation}}
\newcommand{\ee}{\end{equation}}
\newcommand{\bea}{\begin{eqnarray}}
\newcommand{\eea}{\end{eqnarray}}
\newcommand{\beq}{\begin{equation}}
\newcommand{\eeq}{\end{equation}}
\newcommand{\beqn}{\begin{eqnarray}}
\newcommand{\tr}{\\ \nonumber}
\newcommand{\eeqn}{\end{eqnarray}}

\usepackage{amsfonts}
\usepackage{amssymb}
\usepackage{amsmath}
\usepackage{hyperref}
\usepackage{slashed}
\begin{document}
%\preprint{MCTP-14-30}

\def \l {\lambda}
\def \tr {\nonumber\\}
\def \bege {\begin{equation}}
\def \ende {\end{equation}}
\def \beges {\begin{eqnarray}}
\def \endes {\end{eqnarray}}
\def \l {\lambda}
\def \t {\theta}
\def	\k  {\kappa}
\def \e {\epsilon}
\def \le {\leqslant}
\def \sl {\sqrt{\l}}

\title{{\bf One-loop Structure of Higher Rank Wilson Loops in AdS/CFT}}

\author{Alberto Faraggi}\email{faraggi@fis.puc.cl}
\affiliation{Instituto de F\'isica, Pontificia Universidad Cat\'olica de Chile, Casilla 306, Santiago, Chile}

\author{James T. Liu }\email{jimliu@umich.edu}
\affiliation{Michigan Center for Theoretical Physics, University of Michigan, Ann Arbor, MI 48109, USA}

\author{Leopoldo A. Pando Zayas}\email{lpandoz@umich.edu}
\affiliation{Michigan Center for Theoretical Physics, University of Michigan, Ann Arbor, MI 48109, USA}

\author{Guojun Zhang}\email{zgj1@mail.ustc.edu.cn}
\affiliation{Department of Modern Physics, University of Science and Technology of China, Hefei, China 230026}

%\date{\today}
\begin{abstract}
The half-supersymmetric Wilson loop in $\mathcal N=4$ SYM is arguably the central non-local operator in the AdS/CFT correspondence. On the field theory side, the vacuum expectation values of Wilson loops in arbitrary representations of $SU(N)$ are captured to all orders in perturbation theory by a Gaussian matrix model. Of prominent interest are the $k$-symmetric and $k$-antisymmetric representations, whose gravitational description is given in terms of D3- and D5-branes, respectively, with fluxes in their world volumes. At leading order in $N$ and $\lambda$ the agreement in both cases is exact. In this note we explore the structure of the next-to-leading order correction in the matrix model and compare with existing string theory calculations. We find agreement in the functional dependence on $k$ but a mismatch in the numerical coefficients.
\end{abstract}

\maketitle

\section{ Introduction}
Wilson loops are non-local operators that play a central role in field theories, serving as order parameters and as generating functions for all local operators. For the ${\cal N}=4$ supersymmetric Yang-Mills theory the most natural Wilson loop one can consider is
\be
\label{Eqn:WL}
W_{\cal R}=\frac{1}{{\rm dim}[{\cal R}]}{\rm Tr}_{\cal R}{\cal P}\exp\left(i\oint_C ds(A_\mu \dot{x}^\mu +i\Phi_I\Theta^I|\dot{x}|)\right),
\ee
where $A_\mu$ and $\Phi^I$ are the gauge and scalar fields, respectively, taking values in a representation ${\cal R}$ of the $SU(N)$ gauge symmetry algebra, while $x^\mu(s)$ parametrizes a curve $C$ in $\mathbb{R}^{4}$ and $\Theta^I$ is a vector in $\mathbb{R}^6$. When the contour $C$ is a circle and $\Theta^2=1$ the loop preserves half of the supersymmetries of the theory.

From the purely field theoretic side a conjecture for the computation of the exact expectation value by means of a Gaussian matrix model was put forward  in \cite{Drukker:2000rr} \cite{Erickson:2000af} and later rigorously proven by Pestun via localization \cite{Pestun:2007rz}. Some explicit calculations using this matrix model were provided for $k$-symmetric and $k$-antisymmetric representations in \cite{Hartnoll:2006is,Yamaguchi:2006tq,Yamaguchi:2007ps}.

Understanding this object in the context of the AdS/CFT correspondence has been an important problem for over a decade. The duality provides an alternative way of computing the vacuum expectation value of (\ref{Eqn:WL}) at strong coupling. The original prescription, proposed in \cite{Maldacena:1998im,Rey:1998ik}, identifies the vev of this Wilson loop in the fundamental representation of $SU(N)$ with the partition function of a fundamental string pinching the loop at the boundary. For higher order representations the string theory origin of the configurations was clarified in terms of D5-branes \cite{Gomis:2006sb} and D3-branes \cite{Gomis:2006im} with $k$-units of flux in their world volumes, corresponding to the $k$-antisymmetric and $k$-symmetric representations, respectively.

The regularized action of the corresponding brane configuration computes the expectation value of the dual Wilson loop at leading order in $N$ and $\lambda$. The calculation was performed for the D3-brane in \cite{Drukker:2005kx} and for the D5-brane in \cite{Pawelczyk:2000hy,Camino:2001at}, finding exact agreement with the matrix model result.

Our purpose in this letter is to summarize the state of affairs at one-loop level. Focusing on the $k$-symmetric and $k$-antisymmetric representations, we discuss certain calculable corrections to the leading order results for the half-supersymmetric circular Wilson loop (\ref{Eqn:WL}) from the field theoretic perspective and compare them with the gravitational predictions. After revealing numerical discrepancies between the two approaches, we highlight particular aspects of the calculations which could be the source of the mismatch. We contrast our results with previous attempts for the fundamental representation \cite{Drukker:2000ep,Sakaguchi:2007ea,Kruczenski:2008zk} and argue that the rank $k$ of the representations provides a new knob that might allow us to establish agreement beyond the leading order.

%Our purpose in this letter is to summarize the state of affairs at one-loop level. We highlight the numerical discrepancies at this level and point out particular aspects of the calculations where the ultimate resolution might be found. Namely, we discuss certain calculable corrections to the leading order results for the half-s%upersymmetric circular Wilson loop (\ref{Eqn:WL}) from the field theoretic perspective and compare to the gravitational predictions. We focus on the case of $k$-symmetric and $k$-antisymmetric representations and contrast our results with previous attempts for the fundamental representation \cite{Drukker:%2000ep,Sakaguchi:2007ea,Kruczenski:2008zk}. We argue that the rank $k$ of the representations provides a new knob that might allow us to establish agreement beyond the leading order.

{\bf Note added.} In the original version of this manuscript we claimed to have found exact agreement between the matrix model calculation for the $k$-symmetric representation and the D3-brane results reported in \cite{Buchbinder:2014nia}. This is incorrect as there is an overall sign difference. The functional dependence on $\kappa$, however, is in perfect agreement.

%%%%%%%%%%%%%%%%%%%%%%%%%%%%%%%%%%%%%%%%%%%%%%%%%%%%%%%%%%%%%%%%%%%%%%%%%%%
%\section{Gauge theory result at one loop}\label{Sec:MatrixModel}
%%%%%%%%%%%%%%%%%%%%%%%%%%%%%%%%%%%%%%%%%%%%%%%%%%%%%%%%%%%%%%%%%%%%%%%%%%%%%%%%
\section{Gauge theory beyond leading order}

The starting point is the Gaussian matrix model defined by the partition function
%
%\begin{empheq}{align}
\be
    Z=\int dM\exp\left(-\frac{2N}{\lambda}\textrm{Tr}\left(M^2\right)\right)\,,
\ee
%\end{empheq}
%
where $M$ is a $N\times N$ matrix and $\lambda$ is the 't Hooft coupling. It is most convenient to work in the eigenvalue basis:  $M={\rm diag}\{m_1, m_2, m_3, \ldots, m_N\}$. Moreover, for the calculation of the expectation values of
(\ref{Eqn:WL}) in the $k$-th symmetric and antisymmetric representations it is useful to consider the generating functions for the relevant polynomials, namely, $F_A(t)=\prod_{i=1}^N (1+te^{m_i})$ and $F_S(t)=\prod_{i=1}^N (1-te^{m_i})^{-1}$ as in \cite{Hartnoll:2006is}. When inserted in the Gaussian matrix model we obtain
\be
\label{Eqn:Master}
\langle F_{A,S}(t)\rangle = \frac{1}{Z}\int \prod\limits_{j=1}^N[dm_j]\Delta^2(m)F_{A,S}\exp\left(-\frac{2N}{\lambda} \sum\limits_{i=1}^N m_i^2\right),
\ee
where $\Delta(m)$ is the Vandermonde determinant coming from the change of integration variables. Up to a normalization, the coefficients of the $t^k$ term in these series yield the expectation values of the Wilson loops in the corresponding representations.

{\bf The large $N$ approximation.} In some cases, it is possible to evaluate (\ref{Eqn:Master}) as an exact expression in $N,\lambda, k$ by using orthogonal polynomials.  A lot of effort has gone into understanding such expressions \cite{Fiol:2013hna}, although the results are somewhat formal and do not highlight
the functional dependence on the parameters.  While we will report progress in this direction in a separate publication, here we will focus on the large $N$ limit.  In this limit, the eigenvalues can be approximated by continuous variables which are well described by the normalized Wigner semi-circle distribution
\be
\rho(m)=\frac{2}{\pi \lambda} \sqrt{\lambda-m^2}, \qquad -\sqrt{\lambda}\le m \le \sqrt{\lambda}.
\label{eq:Wsc}
\ee

\begin{figure}[t]
\begin{tikzpicture}
\draw[thick,->]    (-4,0)--(4,0) node[anchor=south east]{$x$};
\draw[thick,->] (0,-1)--(0,2.3) node[anchor=north west]{$y$};
\draw (-1,0)  sin (-0.9,0.1) cos (-0.8,0) sin (-0.7,-0.1) cos (-0.6,0) sin (-0.5,0.1) cos (-0.4,0) sin (-0.3,-0.1) cos (-0.2,0) sin (-0.1,0.1) cos (0,0) sin (0.1,-0.1) cos (0.2,0) sin (0.3,0.1) cos(0.4,0) sin(0.5,-0.1) cos(0.6,0) sin(0.7,0.1) cos(0.8,0) sin(0.9,-0.1) cos(1,0);
\draw[dashed] (-4,1)--(4,1);
\draw[thick,->] (-2,0)--(-2,0.5);
\draw[thick] (-2,0.5)--(-2,1);
\draw[thick,->] (2,0)--(2,0.5);
\draw[thick] (2,0.5)--(2,1);
\draw[thick,->] (-1,0.3)--(0,0.3);
\draw[thick] (0,0.3)--(1,0.3);
\draw[thick] (1,-0.3) arc (-90:90:0.3);
\draw[thick,->] (1,-0.3)--(0,-0.3);
\draw[thick] (0,-0.3)--(-1,-0.3);
\draw[thick] (-1,0.3) arc (90:270:0.3);
\node at (0.4,1.4) {$\frac{2\pi i}{\sqrt{\lambda}}$};
\node at (-1.8,0.5) {$C$};
\node at (1,-0.6) {$C_{branch}$};
\node at (2.2,0.5) {$C'$};
\end{tikzpicture}
\caption{\label{fig:Cbranch}The contour of integration $C$ for the $k$-symmetric representation, and its deformation
into $C'$ and $C_{\rm branch}$.}
\end{figure}
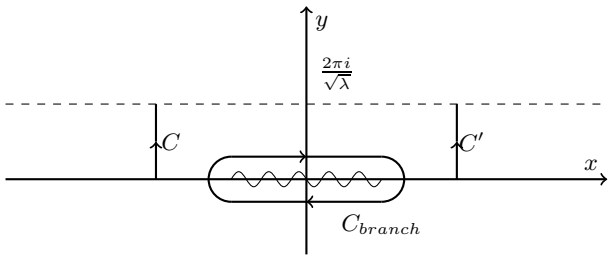

The expectation value of the Wilson loop in the rank-$k$ representation can be obtained from (\ref{Eqn:Master}) by means of
the residue theorem.  Following \cite{Hartnoll:2006is}, we let $f=k/N$ and
make the transformation $t\to e^{\sqrt{\lambda}z}$, which maps the plane to a cylinder, to obtain
\begin{widetext}
\be
\langle W_{S,A}\rangle=d_{S,A}^{-1}\frac{\sqrt{\lambda}}{2\pi i}\int_C dz \exp\left(\mp N\left[\frac{2}{\pi}\int_{-1}^{1} dx \sqrt{1-x^2}\log\left(1\mp e^{\sqrt{\lambda}(-x+z)}\right)
\pm f\sqrt{\l}z\right]\right),
\label{eq:Wsa}
\ee
%\end{widetext}
where $d_S,A$ are the dimensions of the respective representations
\begin{equation}
d_S=\binom{N+k-1}k,\qquad
d_A=\binom{N}k.
\end{equation}
Here we have scaled $\sqrt\lambda$ out of the distribution (\ref{eq:Wsc}). The contour $C$ wraps the cylinder once and is taken to the left of any singularities.

Our goal is to evaluate the contour integral in (\ref{eq:Wsa}) in the limit of large $N$, $\lambda$
and $k$. Since the integrand in the exponent scales like $N$, this
can be performed using Laplace's method or, more generally, the steepest descent method.  While the leading behavior was obtained in \cite{Hartnoll:2006is},
it is straightforward to obtain the next-to-leading order term as well by simply expanding around the saddle point.

%%%%%%%%%%%%%%%%%%%%%%%%%%%%%%%%%%%%%%%%%%%%%%%%%%%%%%%%%%%%%%%%%%%%%%%%%%%%%%%%%%%%%%%%%%%%
\subsection{ Symmetric representation at next-to-leading order}
The $k$-symmetric representation corresponds to the choice of upper signs in (\ref{eq:Wsa}). As argued in \cite{Hartnoll:2006is}, all saddle points lie on the real axis, where the integrand develops a branch cut between $-1$ and $1$. For given $N$ and $k$ there exists a critical value of $\lambda$ for which the saddle point hits the branch cut and moves to the second Riemann sheet. To avoid this complication we deform the contour $C$ by pulling it to the right of the
branch cut, thus enclosing the saddle point, as shown in Fig.~\ref{fig:Cbranch}. Since the integral over $C'$ vanishes in
the limit $\mathrm{Re}\;z\to\infty$, we are left with the jump across the cut \cite{Hartnoll:2006is}
%\begin{widetext}
\be
\langle W_S\rangle=d_S^{-1}\frac{\sl}{\pi}\mathrm{Im}\int_{-1}^1 dy \exp\left[-N\left(\frac{2}{\pi}\int_{-1}^1 dx\sqrt{1-x^2}
\log\left(e^{\sl x}-e^{\sl y}\right)+4i\int_{-1}^y dx\sqrt{1-x^2}+f\sl y\right)\right].
\label{eq:Ws}
\ee
%\end{widetext}
For $N\to\infty$ the $y$ integral in (\ref{eq:Ws}) can be evaluated using steepest descent. In the large $\l$ but fixed $\kappa = k\sqrt{\lambda}/(4N)$ limit, it is dominated by the saddle point at $y_0=-\sqrt{1+\k^2}<-1$, which renders the leading contribution. Taking \linebreak
\end{widetext}
into account the pre-factor of $\sqrt{\lambda}=4\kappa/f$ in (\ref{eq:Ws}), we find that the asymptotic result for the $k$-symmetric Wilson loop expectation value is

\bea
\label{Eq:MMSym}
\langle W_{S_k}\rangle &=& \exp\biggl(2N\left[\kappa\sqrt{1+\kappa^2}+\sinh^{-1}\kappa\right] \nonumber \\
&&\qquad+\frac{1}{2}\ln \frac{\kappa^3}{\sqrt{1+\kappa^2}} \biggr).
\eea

%%%%%
%%%%%

%%%%%%%%%%%%%%%%%%%%%%%%%%%%%%%%%%%%%%%%%%%%%%%%%%%%%%%%%%%%5
\subsection{Antisymmetric representation at next-to-leading order}
The antisymmetric case, which corresponds to choosing the lower signs in (\ref{eq:Wsa}), is simpler to analyze. As shown in \cite{Hartnoll:2006is}, the saddle point lies on the real axis, whereas the branch cut displayed by the integrand has an imaginary part. This allows us to directly calculate the saddle point from (\ref{eq:Wsa}). Deforming the contour appropriately and taking the large $\l$ limit and $N\to\infty$ while keeping $k/N$ fixed, we find
%Taking the first saddle point correction to (\ref{Eqn:Master}), we find that ($N\to \infty, k/N$-fixed)
\be
\langle W_{A_k}\rangle = \exp\left(\frac{2N\sqrt{\lambda}}{3\pi}\sin^3\theta_k-\frac{1}{2}\ln \sin\theta_k \right),
\label{eq:Wak}
\ee
where $\theta_k$ is given by the solution to $k=N(\theta_k-\sin \theta_k \cos \theta_k  )/\pi$. Notice that the leading term in the exponent in (\ref{eq:Wak}) is proportional to $N\sqrt{\lambda}$ when $\lambda$ is large, in contrast to the symmetric case where (\ref{Eq:MMSym}) goes like $N$. As we will see below, this is consistent with the corrections in the gravitational description.

We should mention that the integrand in (\ref{eq:Wsa}) exhibits a second saddle point that lies on top of the branch cut. We have verified that by deforming the contour to wrap the branch cut, computing the discontinuity across it, and evaluating the resulting integral using the saddle point approximation, yields the same result at the leading and sub-leading levels as above.

\section{Gravity results: D3 and D5 branes with worldvolume flux in $AdS_5\times S^5$}
%%%%%%%%%%%%%%%%%%%%%%%%%%%%%%%%%%%%%%%%%%%%%%%%%%%%%%%%%%%%%%%%%%%%%%%%%%%%%%%%%%%%%%%%%%%%%%

According to the holographic dictionary, the expectation value of the Wilson loop at leading order is given by the regularized on-shell action of the corresponding dual string theoretic object. For the case of the circular Wilson loop in the fundamental representation this object is a fundamental string with AdS$_2$ worldvolume. This operator has been studied for over a decade now and there has been a concerted effort in trying to match the field theory result with the string theory calculation at higher orders, starting with the insightful works \cite{Forste:1999qn} and \cite{Drukker:2000ep} and more recently in \cite{Sakaguchi:2007ea,Kruczenski:2008zk,Buchbinder:2014nia}.  {\it It is fair to say that the current state is that the two sides do not seem to coincide for the supersymmetric Wilson loop in the fundamental representation}. A number of reasons have been advanced for this mismatch, including the role of zero modes in string disc amplitudes \cite{Drukker:2000ep,Kruczenski:2008zk,Buchbinder:2014nia} .

The situation is slightly better for the supersymmetric Wilson loop in higher representations. The rank of the representation $k$ can scale with $N$, introducing a natural parameter that acts as a new knob in the problem. Indeed, in the cases of the totally symmetric and totally antisymmetric rank $k$ representations we find agreement of the next-to-leading correction to the expectation value of the Wilson loop the at the functional level (as a function of a parameter related to $k$).

For the higher rank representations, the appropriate gravity configuration is given by D$p$-branes, whose bosonic action in AdS$_5\times S^5$ is
\begin{equation}
\label{Eqn:Daction}
	S_{\mathrm Dp}
	     = - T_p \int d^{p+1} \xi \sqrt{-\det(g+{\cal F})_{ab}} + T_p \int C_{(4)}\wedge e^{\cal F}.
\end{equation}
Here $T_d$ is the brane tension. The classical actions yield the expectation value of the Wilson loop at leading order.

{\bf D3-brane.} The classical configuration corresponding to the half supersymmetric Wilson loop in the rank $k$ symmetric representation is a D3-brane embedded in $AdS_5\times S^5$ (with radii $L$) that sits at a fixed point in the $S^5$ part of the background, while spanning an AdS$_2\times S^2 \subset AdS_5$ world volume with radii $L\cosh(u_k)$ and $L\sinh(u_k)$, respectively \cite{Drukker:2005kx}. The parameter $u_k$ is related to the string charge $k$ by $\sinh(u_0)=k\sqrt{N}/4N\equiv\kappa$. This solution is supported by a Euclidean gauge field strength $2\pi \alpha' F=iL^2 \cosh(u_k)$ along AdS$_2$.

The question of quantum corrections to the classical action was boosted by the step taken in \cite{faraggi:2011bb} which organized the spectrum of excitations of all the supergravity objects dual to the circular Wilson loop into supermultiplets of $OSp(4^*|4)$. This work sets the stage for the calculation of the one loop effective actions in the gravity side.

As shown in \cite{faraggi:2011bb}, the quadratic fluctuations around the above solution take the form
\bea
S_{B}&=& \frac{T_{\mathrm D3}\coth(u_k)}{2}\int d^4\sigma \sqrt{g}\bigg[g^{\alpha\beta}\partial_\alpha \phi^i\partial_\beta \phi^i \nonumber\\
&&\kern10em+g^{\alpha\beta}g^{\gamma\delta}f_{\alpha \gamma} f_{\beta \delta}\bigg], \nonumber \\
S_{F}&=&T_{\mathrm D3}\coth(u_k)\int d^4\sigma \sqrt{g}\bar{\Theta}_A\slashed{\nabla}\Theta_A.
\eea
As appropriate to a D3-brane, the field content is that of an $\mathcal{N}=4$ vector multiplet: six scalars $\phi^i, i=1, \ldots 6$, a gauge field $a_{\alpha}$, with field strength $f_{\alpha\beta}$, and a Weyl spinor transforming in the {\bf 4} of $SO(6) \simeq SU(4)$. The fact that all fields are massless is a consequence of the supersymmetry preserved by the background. The $AdS_2\times S^2$ geometry is not precisely the induced one but the so-called open string metric for which both $AdS_2$ and $S^2$ have the same radius $L\sinh(u_k)$. Notice that $T_{\mathrm D3}L^4 \sim N$, implying that the expansion parameter is  $1/N$, in accordance with the dual description.

The one-loop effective action was recently computed in \cite{Buchbinder:2014nia} using heat kernel techniques \cite{Vassilevich:2003xt} thoroughly explained in \cite{Banerjee:2010qc,Banerjee:2011jp} for the case of fields in $AdS_2 \times S^2$ in the context of logarithmic corrections to black hole entropy. As it turns out, the normalization factor of $\coth(u_k)$ is crucial in the calculation as it contributes to the one-loop effective action in a non-trivial ($\kappa$--dependent) fashion. The result is
\bea\label{eq:WlsD3}
\int\exp\left(-S_{\mathrm D3}\right)&=& \exp\biggl(2N\left[\kappa\sqrt{1+\kappa^2}+\sinh^{-1}\kappa\right]\nonumber \\
&&\qquad- \frac{1}{2}\ln \frac{\kappa^3}{\sqrt{1+\kappa^2}} \biggr)\,.
\eea
The leading/classical term was known to match the field theory calculation \cite{Drukker:2005kx,Hartnoll:2006is}. The second line matches the matrix model calculation given in  Eq.~(\ref{Eq:MMSym}), except for an overall sign. %As noted by Buchbinder-Tseyltin \cite{Buchbinder:2014nia}, in the small-$\kappa$ limit this answer matches the exact $1/N$ correction to the $\Box^k$ representation,  Eq. (\ref{kFundExact}), which is to be identified with a $k$-wound string. This arguments speaks of the robustness of the negative sign in the above expression. %This analysis potentially point to the robustness of the negative sign here%

{\bf D5-brane.} In the case of the Wilson loop in the totally antisymmetric rank $k$ representation, the dual object is a D5-brane whose classical solution has an AdS$_2\times S^4$ world volume with radii $L$ and $L\sin\theta_k$, respectively \cite{Camino:2001at,Yamaguchi:2006tq}.  The string charge is related to $\theta_k $  by $k=N(\theta_k-\sin \theta_k \cos \theta_k  )/\pi$. Excitations of the brane correspond to fields propagating on this space, albeit with an open string metric with a common radius  $L\sin\theta_k$.  The quadratic action is given by \cite{Faraggi:2011ge}
\bea
\label{Eq:D5Action}
S_{B}&=&\frac{T_{\mathrm D5}}{2 \sin\theta_k}\int d^6\xi \sqrt{g}\bigg[\chi^i (\nabla_a \nabla^a -\frac{2}{L^2})\chi_i \nonumber \\
&&\qquad+ \chi^5(\nabla_a \nabla^a +\frac{4}{L^2})\chi_5 -\frac{1}{2}f^{\mu\nu}f_{\mu\nu} -f^{\mu\alpha}f_{\mu\alpha}\nonumber\\
&&\qquad-\frac{1}{2}f^{\alpha\beta}f_{\alpha\beta}
-\frac{4i}{L}\chi^5\epsilon^{\alpha\beta}f_{\alpha\beta}\bigg] ,\nonumber \\
S_{F}&=&\frac{T_{\mathrm D5}}{2 \sin\theta_k}\int d^6\xi\sqrt{g}\,\bar{\Theta}\bigg[\Gamma^a\nabla_a +\frac{1}{L}\Gamma^{6789}\bigg]\Theta,
\eea
where $\alpha,\beta$ are indices in AdS$_2$ and $\mu, \nu$ are indices on $S^4$. The index $i=1,2,3$ denotes the three fluctuations of the embedding AdS$_2\subset \mathrm{AdS}_5$ while the field $\chi^5$ represents the fluctuations of the embedding $S^4 \subset S^5$; thus we have a total of four scalar fields. The multiplet also contains a gauge field with field strength $f$  and a 6d symplectic Majorana spinor $\Theta$. In this case, the quantum corrections are controlled by the parameter $T_{D5}L^6\sim N\sqrt{\lambda}$. The same is true in the matrix model depiction of the operator.

It is worth pointing out that the final term in the bosonic action is non-covariant from the 6d point of view; it emerges from the fact that the background field strength is non-vanishing in the $AdS_2$ directions. Despite its appearance, the action (\ref{Eq:D5Action}) is supersymmetric, as was proven explicitly in \cite{Faraggi:2011ge}.

The computation of the one-loop effective action of the D5 dual to the $k$-antisymmetric representation following from the quadratic action  (\ref{Eq:D5Action}) was performed in \cite{Faraggi:2011ge} using heat kernel techniques. Since the action contains a non-covariant term, the most efficient way goes through compactification of the quadratic Lagrangian on $S^4$. The one-loop effective action was found to be
\begin{equation}
\int\exp\left(-S_{\mathrm D5}\right)=\exp\left(\frac{2N\sqrt{\lambda}}{3\pi} \sin^3\theta_k - \frac{1}{6} \ln\sin\theta_k\right).
\label{eq:WlaD5}
\end{equation}
In this result there is a factor of  two difference with respect to the answer reported in \cite{Faraggi:2011ge}; this stems from the fact that here we correct, following a lucid explanation in \cite{Buchbinder:2014nia}, by a contribution due to the normalization of the quadratic modes in the open string frame versus the closed string frame. The leading term matches exactly the gauge theory calculation \citep{Hartnoll:2006is}\cite{Yamaguchi:2006tq}. The next to leading term has the same functional dependence on $\theta_k$ as the matrix model answer but differs by a factor of $3$.

%%%%%%%%%%%%%%%%%%%%%%%%%%%%%%%%%%%%%%%%%%%%%%%%%%%%%%%%%%%%%%%%%%%%%%%%%%%%%%%%%%%%%%%%%%%%%%%%
\section{Conclusions}
In this article we have discussed the half supersymmetric Wilson loop in the context of the AdS/CFT correspondence, with a special emphasis on higher rank representations of $SU(N)$. In particular, we have computed, using the Gaussian matrix model, a $1/N$ correction to the expectation value of this operator in the $k$-symmetric and $k$-antisymmetric representations. Upon comparison with analogous calculations on the gravity side, which consider one-loop corrections around the corresponding classical D3- and D5-brane solutions, we have found that there is functional agreement on the rank of the representation, $k$, but discrepancies in the numerical coefficients: in the case of the $k$-symmetric representation they disagree by an overall sign, while for the $k$-antisymmetric there is disagreement by a factor of $3$.

Having compared the calculations on the two sides of the duality, a few comments about the mismatch are in order. As emphasized by Buchbinder-Tseyltin \cite{Buchbinder:2014nia}, when $\kappa<<1$ the 1-loop correction to the D3-brane effective action (and the gauge theory correction) should approach the first $1/N$ correction to the expectation value of the Wilson loop in the $\Box^k$ representation. This property holds in (\ref{eq:WlsD3}). Indeed, starting from the exact matrix model result
\bea
\langle W_{\Box^k}\rangle &=&\frac{1}{N}e^{k^2\lambda/8N}L_{N-1}^{(1)}(-k^2\lambda/4N),
\eea
either by exploiting the differential equation satisfied by the Laguerre polynomials \cite{Drukker:2005kx,Kawamoto:2008gp,Buchbinder:2014nia} or by simply using their asymptotic expansion, the leading and sub-leading terms are found to be
\bea
\label{kFundExact}
\langle W_{\Box^k}\rangle &=&\exp\left(2N[\kappa\sqrt{1+\kappa^2}]+\sinh^{-1}\kappa] \right.\nonumber \\
&-&\frac{1}{2}\left.\ln\kappa^3\sqrt{1+\kappa^2}\right),
\eea
when $N\rightarrow\infty$ and $\kappa = k\sqrt{\lambda}/(4N)$ is fixed. This argument speaks to the robustness of the negative sign in expression (\ref{eq:WlsD3}) as opposed to the plus sign (\ref{Eq:MMSym}).

A similar argument holds for the $k$-antisymmetric representation, even though a different limit is taken in this case, namely, large $N$ with $k/N$ fixed and large $\lambda$. When $k/N$ is small the correction should approach that of the fundamental representation. Our result (\ref{eq:Wak}) does not comply with this requirement. Neither does the D5-brane calculation (\ref{eq:WlaD5}), however, so the situation is less clear in this case.

The above analysis seems to render our calculations (\ref{Eq:MMSym}) and (\ref{eq:Wak}) in the matrix model framework invalid. If one insists, however, on computing $\langle W_{A_k,S_k}\rangle$ using the generating functions $F_{A,S}(t)=\textrm{det}\left(1\pm te^M\right)^{\pm}$, the type of asymptotic corrections addressed in this article are essentially unavoidable. This is simply a consequence of using the steepest descent method, and is further evidenced by the fact that we do find agreement with the string theory predictions at a functional level. The complete story, of course, must take into account other $1/N$ corrections of different origin. One obvious such correction would come from a more accurate approximation for the eigenvalue distribution of the Gaussian matrix model, beyond the Wigner semi-circle law. Also, in the limit of a continuous density of eigenvalues, the pole structure of the Cauchy integral changes drastically, making the analysis of corrections even more complicated. Finally, we point out that in the case of the $k$-antisymmetric representation the expansion parameter seems to be $1/N\sqrt{\lambda}$ and not $1/N$, as suggested by the gravitational description; perhaps additional $1/\sqrt{\lambda}$ corrections in the matrix model must therefore be computed. We postpone these lines of inquiry for the future.

{\bf Acknowledgments.}
This work is  partially supported by Department of Energy under grant DE-SC0007859. A.F. is supported by CONICYT/PAI Apoyo al Retorno 821320022. He gratefully acknowledges the support given by FAPESP N\'umero do Processo 2012/03437-8 as well as his host institution, Universidade de S\~ao Paulo, during his stay in Brasil. Special thanks to Diego Trancanelli.

%\newpage
% If you don't have the corresponding .bst and .bib files, comment the two lines below and copy paste the content of the .bbl file
%there (whoever compiled the %bibliography should send you the .bbl file)
%\bibliographystyle{JHEP}
\bibliography{WLoops-bib}

\begin{thebibliography}{25}
\expandafter\ifx\csname natexlab\endcsname\relax\def\natexlab#1{#1}\fi
\expandafter\ifx\csname bibnamefont\endcsname\relax
  \def\bibnamefont#1{#1}\fi
\expandafter\ifx\csname bibfnamefont\endcsname\relax
  \def\bibfnamefont#1{#1}\fi
\expandafter\ifx\csname citenamefont\endcsname\relax
  \def\citenamefont#1{#1}\fi
\expandafter\ifx\csname url\endcsname\relax
  \def\url#1{\texttt{#1}}\fi
\expandafter\ifx\csname urlprefix\endcsname\relax\def\urlprefix{URL }\fi
\providecommand{\bibinfo}[2]{#2}
\providecommand{\eprint}[2][]{\url{#2}}

\bibitem[{\citenamefont{Drukker and Gross}(2001)}]{Drukker:2000rr}
\bibinfo{author}{\bibfnamefont{N.}~\bibnamefont{Drukker}} \bibnamefont{and}
  \bibinfo{author}{\bibfnamefont{D.~J.} \bibnamefont{Gross}},
  \bibinfo{journal}{J. Math. Phys.} \textbf{\bibinfo{volume}{42}},
  \bibinfo{pages}{2896} (\bibinfo{year}{2001}), \eprint{hep-th/0010274}.

\bibitem[{\citenamefont{Erickson et~al.}(2000)\citenamefont{Erickson, Semenoff,
  and Zarembo}}]{Erickson:2000af}
\bibinfo{author}{\bibfnamefont{J.~K.} \bibnamefont{Erickson}},
  \bibinfo{author}{\bibfnamefont{G.~W.} \bibnamefont{Semenoff}},
  \bibnamefont{and} \bibinfo{author}{\bibfnamefont{K.}~\bibnamefont{Zarembo}},
  \bibinfo{journal}{Nucl. Phys.} \textbf{\bibinfo{volume}{B582}},
  \bibinfo{pages}{155} (\bibinfo{year}{2000}), \eprint{hep-th/0003055}.

\bibitem[{\citenamefont{Pestun}(2007)}]{Pestun:2007rz}
\bibinfo{author}{\bibfnamefont{V.}~\bibnamefont{Pestun}}
  (\bibinfo{year}{2007}), \eprint{0712.2824}.

\bibitem[{\citenamefont{Hartnoll and Kumar}(2006)}]{Hartnoll:2006is}
\bibinfo{author}{\bibfnamefont{S.~A.} \bibnamefont{Hartnoll}} \bibnamefont{and}
  \bibinfo{author}{\bibfnamefont{S.~P.} \bibnamefont{Kumar}},
  \bibinfo{journal}{JHEP} \textbf{\bibinfo{volume}{08}}, \bibinfo{pages}{026}
  (\bibinfo{year}{2006}), \eprint{hep-th/0605027}.

\bibitem[{\citenamefont{Yamaguchi}(2006)}]{Yamaguchi:2006tq}
\bibinfo{author}{\bibfnamefont{S.}~\bibnamefont{Yamaguchi}},
  \bibinfo{journal}{JHEP} \textbf{\bibinfo{volume}{05}}, \bibinfo{pages}{037}
  (\bibinfo{year}{2006}), \eprint{hep-th/0603208}.

\bibitem[{\citenamefont{Yamaguchi}(2007)}]{Yamaguchi:2007ps}
\bibinfo{author}{\bibfnamefont{S.}~\bibnamefont{Yamaguchi}},
  \bibinfo{journal}{JHEP} \textbf{\bibinfo{volume}{06}}, \bibinfo{pages}{073}
  (\bibinfo{year}{2007}), \eprint{hep-th/0701052}.

\bibitem[{\citenamefont{Maldacena}(1998)}]{Maldacena:1998im}
\bibinfo{author}{\bibfnamefont{J.~M.} \bibnamefont{Maldacena}},
  \bibinfo{journal}{Phys. Rev. Lett.} \textbf{\bibinfo{volume}{80}},
  \bibinfo{pages}{4859} (\bibinfo{year}{1998}), \eprint{hep-th/9803002}.

\bibitem[{\citenamefont{Rey and Yee}(2001)}]{Rey:1998ik}
\bibinfo{author}{\bibfnamefont{S.-J.} \bibnamefont{Rey}} \bibnamefont{and}
  \bibinfo{author}{\bibfnamefont{J.-T.} \bibnamefont{Yee}},
  \bibinfo{journal}{Eur. Phys. J.} \textbf{\bibinfo{volume}{C22}},
  \bibinfo{pages}{379} (\bibinfo{year}{2001}), \eprint{hep-th/9803001}.

\bibitem[{\citenamefont{Gomis and Passerini}(2006)}]{Gomis:2006sb}
\bibinfo{author}{\bibfnamefont{J.}~\bibnamefont{Gomis}} \bibnamefont{and}
  \bibinfo{author}{\bibfnamefont{F.}~\bibnamefont{Passerini}},
  \bibinfo{journal}{JHEP} \textbf{\bibinfo{volume}{08}}, \bibinfo{pages}{074}
  (\bibinfo{year}{2006}), \eprint{hep-th/0604007}.

\bibitem[{\citenamefont{Gomis and Passerini}(2007)}]{Gomis:2006im}
\bibinfo{author}{\bibfnamefont{J.}~\bibnamefont{Gomis}} \bibnamefont{and}
  \bibinfo{author}{\bibfnamefont{F.}~\bibnamefont{Passerini}},
  \bibinfo{journal}{JHEP} \textbf{\bibinfo{volume}{01}}, \bibinfo{pages}{097}
  (\bibinfo{year}{2007}), \eprint{hep-th/0612022}.

\bibitem[{\citenamefont{Drukker and Fiol}(2005)}]{Drukker:2005kx}
\bibinfo{author}{\bibfnamefont{N.}~\bibnamefont{Drukker}} \bibnamefont{and}
  \bibinfo{author}{\bibfnamefont{B.}~\bibnamefont{Fiol}},
  \bibinfo{journal}{JHEP} \textbf{\bibinfo{volume}{02}}, \bibinfo{pages}{010}
  (\bibinfo{year}{2005}), \eprint{hep-th/0501109}.

\bibitem[{\citenamefont{Pawelczyk and Rey}(2000)}]{Pawelczyk:2000hy}
\bibinfo{author}{\bibfnamefont{J.}~\bibnamefont{Pawelczyk}} \bibnamefont{and}
  \bibinfo{author}{\bibfnamefont{S.-J.} \bibnamefont{Rey}},
  \bibinfo{journal}{Phys. Lett.} \textbf{\bibinfo{volume}{B493}},
  \bibinfo{pages}{395} (\bibinfo{year}{2000}), \eprint{hep-th/0007154}.

\bibitem[{\citenamefont{Camino et~al.}(2001)\citenamefont{Camino, Paredes, and
  Ramallo}}]{Camino:2001at}
\bibinfo{author}{\bibfnamefont{J.~M.} \bibnamefont{Camino}},
  \bibinfo{author}{\bibfnamefont{A.}~\bibnamefont{Paredes}}, \bibnamefont{and}
  \bibinfo{author}{\bibfnamefont{A.~V.} \bibnamefont{Ramallo}},
  \bibinfo{journal}{JHEP} \textbf{\bibinfo{volume}{05}}, \bibinfo{pages}{011}
  (\bibinfo{year}{2001}), \eprint{hep-th/0104082}.

\bibitem[{\citenamefont{Drukker et~al.}(2000)\citenamefont{Drukker, Gross, and
  Tseytlin}}]{Drukker:2000ep}
\bibinfo{author}{\bibfnamefont{N.}~\bibnamefont{Drukker}},
  \bibinfo{author}{\bibfnamefont{D.~J.} \bibnamefont{Gross}}, \bibnamefont{and}
  \bibinfo{author}{\bibfnamefont{A.~A.} \bibnamefont{Tseytlin}},
  \bibinfo{journal}{JHEP} \textbf{\bibinfo{volume}{04}}, \bibinfo{pages}{021}
  (\bibinfo{year}{2000}), \eprint{hep-th/0001204}.

\bibitem[{\citenamefont{Sakaguchi and Yoshida}(2008)}]{Sakaguchi:2007ea}
\bibinfo{author}{\bibfnamefont{M.}~\bibnamefont{Sakaguchi}} \bibnamefont{and}
  \bibinfo{author}{\bibfnamefont{K.}~\bibnamefont{Yoshida}},
  \bibinfo{journal}{Nucl.Phys.} \textbf{\bibinfo{volume}{B798}},
  \bibinfo{pages}{72} (\bibinfo{year}{2008}), \eprint{0709.4187}.

\bibitem[{\citenamefont{Kruczenski and Tirziu}(2008)}]{Kruczenski:2008zk}
\bibinfo{author}{\bibfnamefont{M.}~\bibnamefont{Kruczenski}} \bibnamefont{and}
  \bibinfo{author}{\bibfnamefont{A.}~\bibnamefont{Tirziu}},
  \bibinfo{journal}{JHEP} \textbf{\bibinfo{volume}{05}}, \bibinfo{pages}{064}
  (\bibinfo{year}{2008}), \eprint{0803.0315}.

\bibitem[{\citenamefont{Buchbinder and Tseytlin}(2014)}]{Buchbinder:2014nia}
\bibinfo{author}{\bibfnamefont{E.}~\bibnamefont{Buchbinder}} \bibnamefont{and}
  \bibinfo{author}{\bibfnamefont{A.}~\bibnamefont{Tseytlin}},
  \bibinfo{journal}{Phys.Rev.} \textbf{\bibinfo{volume}{D89}},
  \bibinfo{pages}{126008} (\bibinfo{year}{2014}), \eprint{1404.4952}.

\bibitem[{\citenamefont{Fiol and Torrents}(2014)}]{Fiol:2013hna}
\bibinfo{author}{\bibfnamefont{B.}~\bibnamefont{Fiol}} \bibnamefont{and}
  \bibinfo{author}{\bibfnamefont{G.}~\bibnamefont{Torrents}},
  \bibinfo{journal}{JHEP} \textbf{\bibinfo{volume}{1401}}, \bibinfo{pages}{020}
  (\bibinfo{year}{2014}), \eprint{1311.2058}.

\bibitem[{\citenamefont{Forste et~al.}(1999)\citenamefont{Forste, Ghoshal, and
  Theisen}}]{Forste:1999qn}
\bibinfo{author}{\bibfnamefont{S.}~\bibnamefont{Forste}},
  \bibinfo{author}{\bibfnamefont{D.}~\bibnamefont{Ghoshal}}, \bibnamefont{and}
  \bibinfo{author}{\bibfnamefont{S.}~\bibnamefont{Theisen}},
  \bibinfo{journal}{JHEP} \textbf{\bibinfo{volume}{9908}}, \bibinfo{pages}{013}
  (\bibinfo{year}{1999}), \eprint{hep-th/9903042}.

\bibitem[{\citenamefont{Faraggi and Pando~Zayas}(2011)}]{faraggi:2011bb}
\bibinfo{author}{\bibfnamefont{A.}~\bibnamefont{Faraggi}} \bibnamefont{and}
  \bibinfo{author}{\bibfnamefont{L.~A.} \bibnamefont{Pando~Zayas}},
  \bibinfo{journal}{JHEP} \textbf{\bibinfo{volume}{05}}, \bibinfo{pages}{018}
  (\bibinfo{year}{2011}), \eprint{1101.5145}.

\bibitem[{\citenamefont{Vassilevich}(2003)}]{Vassilevich:2003xt}
\bibinfo{author}{\bibfnamefont{D.}~\bibnamefont{Vassilevich}},
  \bibinfo{journal}{Phys.Rept.} \textbf{\bibinfo{volume}{388}},
  \bibinfo{pages}{279} (\bibinfo{year}{2003}), \eprint{hep-th/0306138}.

\bibitem[{\citenamefont{Banerjee
  et~al.}(2011{\natexlab{a}})\citenamefont{Banerjee, Gupta, and
  Sen}}]{Banerjee:2010qc}
\bibinfo{author}{\bibfnamefont{S.}~\bibnamefont{Banerjee}},
  \bibinfo{author}{\bibfnamefont{R.~K.} \bibnamefont{Gupta}}, \bibnamefont{and}
  \bibinfo{author}{\bibfnamefont{A.}~\bibnamefont{Sen}},
  \bibinfo{journal}{JHEP} \textbf{\bibinfo{volume}{1103}}, \bibinfo{pages}{147}
  (\bibinfo{year}{2011}{\natexlab{a}}), \eprint{1005.3044}.

\bibitem[{\citenamefont{Banerjee
  et~al.}(2011{\natexlab{b}})\citenamefont{Banerjee, Gupta, Mandal, and
  Sen}}]{Banerjee:2011jp}
\bibinfo{author}{\bibfnamefont{S.}~\bibnamefont{Banerjee}},
  \bibinfo{author}{\bibfnamefont{R.~K.} \bibnamefont{Gupta}},
  \bibinfo{author}{\bibfnamefont{I.}~\bibnamefont{Mandal}}, \bibnamefont{and}
  \bibinfo{author}{\bibfnamefont{A.}~\bibnamefont{Sen}},
  \bibinfo{journal}{JHEP} \textbf{\bibinfo{volume}{1111}}, \bibinfo{pages}{143}
  (\bibinfo{year}{2011}{\natexlab{b}}), \eprint{1106.0080}.

\bibitem[{\citenamefont{Faraggi et~al.}(2012)\citenamefont{Faraggi, Mueck, and
  Pando~Zayas}}]{Faraggi:2011ge}
\bibinfo{author}{\bibfnamefont{A.}~\bibnamefont{Faraggi}},
  \bibinfo{author}{\bibfnamefont{W.}~\bibnamefont{Mueck}}, \bibnamefont{and}
  \bibinfo{author}{\bibfnamefont{L.~A.} \bibnamefont{Pando~Zayas}},
  \bibinfo{journal}{Phys.Rev.} \textbf{\bibinfo{volume}{D85}},
  \bibinfo{pages}{106015} (\bibinfo{year}{2012}), \eprint{1112.5028}.

\bibitem[{\citenamefont{Kawamoto et~al.}(2009)\citenamefont{Kawamoto, Kuroki,
  and Miwa}}]{Kawamoto:2008gp}
\bibinfo{author}{\bibfnamefont{S.}~\bibnamefont{Kawamoto}},
  \bibinfo{author}{\bibfnamefont{T.}~\bibnamefont{Kuroki}}, \bibnamefont{and}
  \bibinfo{author}{\bibfnamefont{A.}~\bibnamefont{Miwa}},
  \bibinfo{journal}{Phys.Rev.} \textbf{\bibinfo{volume}{D79}},
  \bibinfo{pages}{126010} (\bibinfo{year}{2009}), \eprint{0812.4229}.

\end{thebibliography}

\end{document}